# Peculiarities of charged particle kinetics in spherical plasma

V I Kolobov[1,2], R R Arslanbekov[2] and H Liang[1]

[1]The University of Alabama in Huntsville, Huntsville, AL 35899, USA
[2]CFD Research Corporation, Huntsville, AL 35806, USA

vladimir.kolobov@uah.edu

**Abstract.** We describe kinetic simulations of transient problems in partially ionized weakly-collisional plasma around spherical bodies absorbing or emitting charged particles. Numerical solutions of kinetic equations for electrons and ions in 1D2V phase space are coupled to an electrostatic solver using the Poisson equation or quasineutrality condition for small Debye lengths. The formation of particle groups and their contributions to electric current flow and screening of charged bodies by plasma are discussed for applications to Langmuir probes and solar wind.

## 1. Introduction

Problems associated with screening spherical bodies in plasmas (Langmuir probes, dust particles, satellites in the ionosphere, solar wind, etc.) are similar. In all these problems, one type of charged particle gets trapped near an electrically charged body by the electric fields, which are induced by the body and self-generated by the surrounding plasma. Due to plasma screening, the trapped particles modify the electric field profile around the body compared to the Coulomb profile (in a vacuum). Dust particles in plasma are charged negatively, and the trapped particles are positive ions. Sun is charged positively, and the trapped particles are electrons. In the case of Langmuir probes, both situations are possible. The floating potential of the probe corresponds to a negatively charged probe, ensuring net zero current to the surface and ion trapping around the probe.

Thermo- and photo-electron emission often occurs from probes and dust particles. Sun emits electrons and ions, and their fluxes must be balanced in the dynamic quasi-steady state of ambient solar wind. Kinetic equations for collisionless non-magnetized plasma have striking similarities with those for magnetized solar wind plasma, especially for the magnetic field of a magnetic monopole. However, we note an essential difference between unmagnetized and magnetized spherical plasma. In the first case, trapped particles circle the sphere on closed orbits around the force center. Trapped particles bounce radially along the magnetic field lines in the second case.

The kinetic theory of plasma around satellites [1] and the kinetic theory of ambient solar wind [2] have been developed. However, their implementation for practical simulations of probes under different conditions is still missing, and relations between the kinetic and commonly used MHD models for the solar wind are unclear. Effects of collisions on particle trapping and screening of charged objects in plasma remain a subject of active research [3,4]. Several recent papers have emphasized the importance of the electric fields in the solar wind for explaining the measured electron velocity distribution functions [5,6], which usually contain three electron groups: core, strahl (beam), and halo. The latest paper [7] explains that the halo could form from the strahl without scattering due to electron reflections by the ambipolar electric field.

This paper describes a kinetic model of spherical plasma coupling grid-based kinetic solvers for electrons and ions in phase space with electrostatic solvers for the electric field. We use a Poisson and an iterative solver for the electric field based on quasineutrality conditions. The latter is crucial for the solar wind due to the tiny value of the Debye length close to the Sun compared to the spatial scale of interest. We illustrate the current capabilities of our model for applications to Langmuir probe and solar wind. We also introduce a quasi-diffusion kinetic model, which could be a part of a future hybrid code for more efficient treatment of different electron groups.

## 2. Kinetic model of spherical plasmas

Assuming spherical symmetry, we reduce the general 3D3V phase space to a 1D2V phase space, which consists of 1D physical space (the radial distance $r$) and 2D velocity space (the magnitude of velocity $v$ and the cosine of the angle between the velocity and the radial direction $\mu$). The kinetic equation in the conservative form has the form [8]:

$$\frac{\partial Y}{\partial t} + \frac{\partial}{\partial r}(\mu v Y) + \frac{\partial}{\partial \mu}\left\{\left[\frac{v(1-\mu^2)}{r} + \frac{qE}{m}\frac{1-\mu^2}{v}\right]Y\right\} + \frac{\partial}{\partial v}\left[\left(\frac{qE}{m}\mu\right)Y\right] = S(r,v,\mu), \quad (1)$$

where $Y = r^2 v^2 f$, $f = f(r,v,\mu)$ is the phase space density, $q$ and $m$ are the charge and mass of the given species, respectively, $E$ is the radial electric field, and $S(r,v,\mu)$ is a collision operator that can describe different types of collisions and wave-particle interactions [9]. The electrostatic potential $\phi$ or the potential energy $U = \frac{e\phi}{k_B T_e}$ normalized to electron thermal energy is usually found from the Poisson equation:

$$\nabla^2 \phi = \frac{e}{\varepsilon_0}(n_e - n_i) \Rightarrow \tilde{\nabla}^2 U = \frac{r_0^2}{\lambda_{De}^2}(\tilde{n}_e - \tilde{n}_i) \quad (2)$$

where, $n_e$ and $n_i$ are electron and ion density, $\tilde{n}_e$ and $\tilde{n}_i$ are the densities normalized to a typical density $n_0$. We normalize the kinetic equation with a typical length $r_0$, typical velocity $v_0$ (~ to the thermal speed $v_{th}$), and typical time scale $t_0$, and use the Poisson equation for the normalization of the electric field $E_0 = 4\pi e n_0 r_0$, where the variables with a tilde are the normalized quantities.

There are two challenges. The value of $\tilde{\lambda}_D^2$ becomes very small at $r_0 " \lambda_D$. This leads to severe inefficiency in calculating the electric field using the Poisson solver and requires another method for calculating the electric field. Another challenge is associated with the electron and ion mass disparity and the speed of electrons and ions. This disparity makes our explicit kinetic solvers very expensive during the transient process (when $\tilde{n}_e \neq \tilde{n}_i$) due to CFL limitations. This challenge can be resolved using an artificially smaller mass ratio, which does not substantially affect the steady solution [10]. The method is based on the fact that in steady-state collisionless electrostatic plasma, the solution with a reduced mass ratio can be scaled to the one for the actual mass ratio. It has been successfully used for speed-limited particle-in-cell (SLPIC) simulations [10].

We solve the kinetic equations for electrons and ions using the same mesh in phase space, which can be dynamically adapted as described in [8]. The Poisson solver uses separate mesh and the Thomas algorithm to calculate the self-consistent electric field based on the number density of electrons and ions obtained from the phase space density. Below, we describe an iterative method of calculating electric fields for problems with small Debye lengths. This coupled kinetic solver for charged particles and the electric field applies to many spherical plasma problems. Below, we demonstrate its current capabilities for simulations of Langmuir probes and solar wind. We also introduce reduced kinetic models for trapped particles.

## 3. Langmuir Probes

Simulations of spherical probes were conducted for $\frac{T_i}{T_e} = 1$ with different $m_r = \frac{m_i}{m_e}$, and $\lambda_{De}$. Figure 1 shows the results for $m_r = 4$ for different $\lambda_{De}$. The electric potential profile $\tilde{\phi}(\tilde{r})$ is calculated self-consistently by the Poisson solver. The flux is reasonably consistent with the analytical predictions. The size of the space charge sheath is proportional to the $\tilde{\lambda}_{De}$ value. When $\tilde{\lambda}_{De} = 0.02$, the potential shows fluctuations near the left boundary, which is unphysical and indicates problems with the Poisson solver.

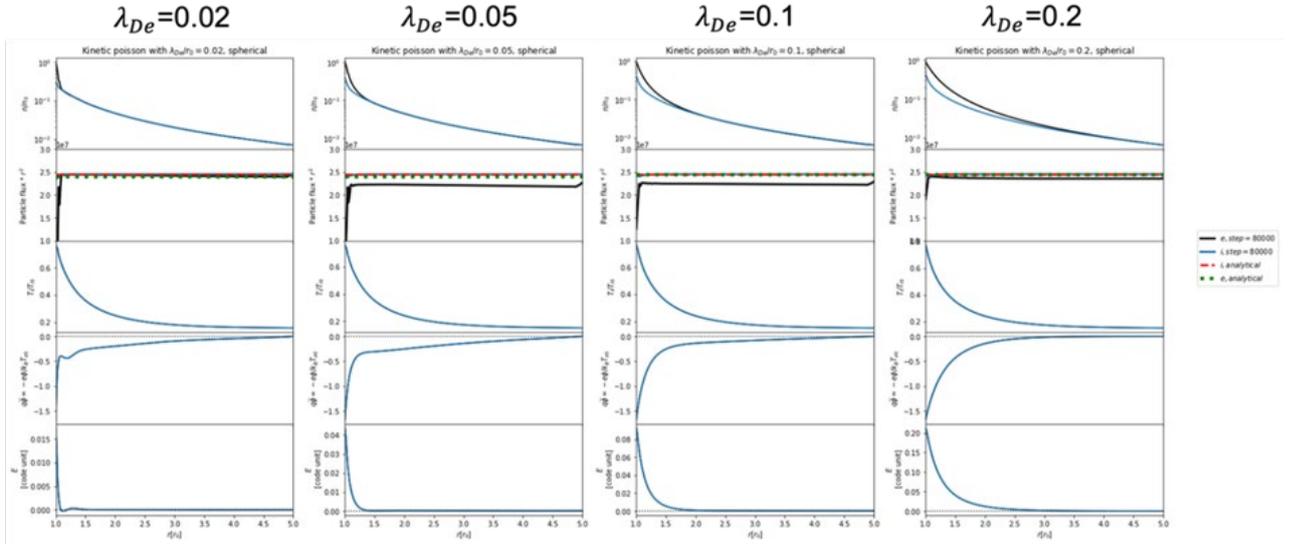

Figure 1 The profiles of density, the particle flux multiplied by $r^2$, temperature, electron potential energy $-\frac{\tilde{\phi}(\tilde{r})}{\tilde{\lambda}_{De}^2}$, and electric field for the mass ratio $m_r = 4$ and $\tilde{\lambda}_{De} = 0.02, 0.05, 0.1,$ and $0.2$. The black (blue) solid line is for electron (ion) profiles. For the particle flux, the green (red) dashed line shows the analytical solution for electrons.

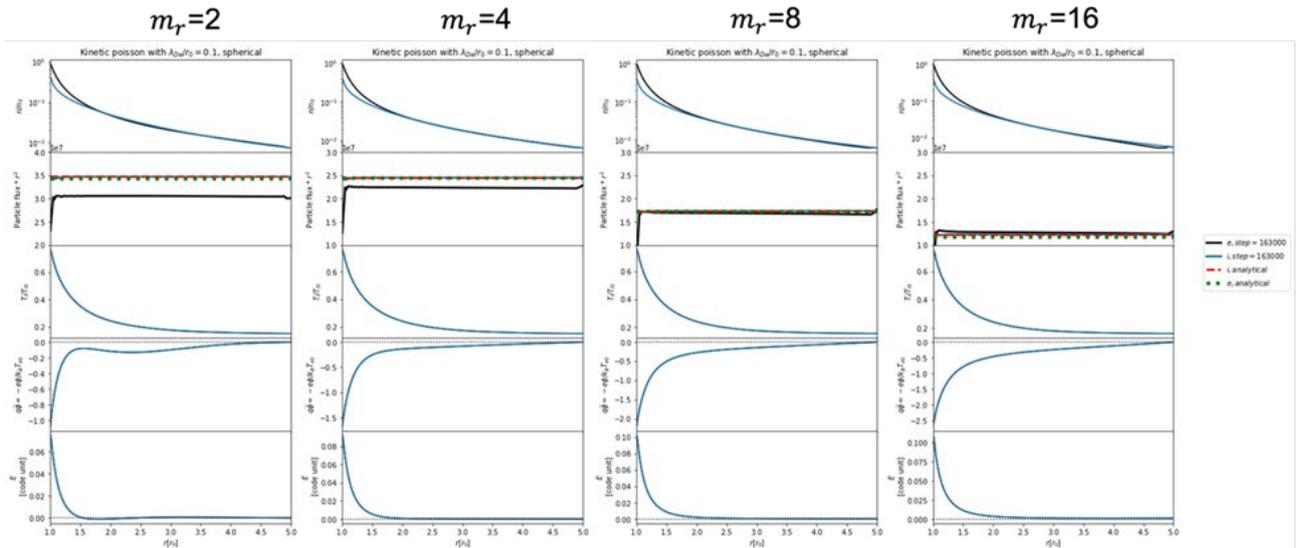

Figure 2 The profiles of density, the particle flux multiplied by $r^2$, temperature, electron potential energy $-\frac{\tilde{\phi}(\tilde{r})}{\tilde{\lambda}_{De}^2}$, and the electric field for fixed $\tilde{\lambda}_{De} = 0.1$ and mass ratio $m_r = 2, 4, 8$ and $16$. The black (blue) solid line is for electron (ion) profiles. For the particle flux, the green (red) dashed line shows the analytical solution for electrons (ions).

Figure 2 shows the results with $\tilde{\lambda}_{De} = 0.1$ for different $m_r$. The calculated $\frac{\tilde{\phi}_L}{\tilde{\lambda}_{De}^2}$ is different for each case. The larger $m_r$, the easier for the system to reach equilibrium. The flux is reasonably consistent with the analytical prediction for $m_r = 4, 8, 16$. The location $\tilde{r}_c$ where the potential nearly reaches its asymptotic behavior increases with increasing $m_r$. This is consistent with the kinetic theory of solar wind predicting $\tilde{r}_c \approx (m_i/m_e)^{1/4}$ [11]. Most of the electrons at $r < \tilde{r}_c$ are returning back to Sun, while the electrons at $r > \tilde{r}_c$ include escaping electrons and those trapped on both sides due to the electric and magnetic mirror forces [11]. The trapped electron trajectories are filled up in a steady state via collisions. In our transient simulations, trapped electrons can also be generated without collisions during the transient effects [12].

## 4. Kinetic Model of Solar Wind

The kinetic equation for charged particles in magnetized plasma coincides with Eq. (1) for a magnetic monopole [13]. In collisionless (exospheric) models, an analytical solution of the kinetic equations for electrons and ions can be obtained using a Liouville theorem. For a Maxwellian distribution of injected particles at the boundary, the analytical solution gives for the electron and ion fluxes [14,11]. In our code, the potential on the right boundary $\phi_R$ was set to zero, and the potential on the left boundary, $\phi_L$ was determined to equalize the ion and electron fluxes in a steady state. The calculated $\phi_L$ was used as a boundary condition for the Poisson solver to find $\phi(r)$ at different $\lambda_{De}$.

This method of calculating the potential drop, $\Delta\varphi^{nocol}$, ensuring equal fluxes of electrons and ions assumes a collisionless plasma and does not consider numerical diffusion, transients, and other factors. Therefore, our code has implemented a more general method based on an interactive potential adjustment to equalize the electron and ion particle fluxes. The scheme uses an adjustable relaxation parameter $\alpha$ (which is proportional to the sphere capacitance), and the analytical value of $\Delta\varphi^{nocol}$ as an initial condition:

$$\Delta\varphi^{n+1} = \Delta\varphi^n + \alpha \Delta\varphi^{nocol} \frac{j_e^n - j_i^n}{|j_e^n + j_i^n|} \quad (3)$$

with $\Delta\varphi^{n=0} = \Delta\varphi^{nocol}$. This scheme allowed converging towards a solution with equal electron and ion particle fluxes.

We first obtained 1D2V solutions assuming constant ion velocity ($v_i = const$) and the ion density profile $n_i(r) = n_0(R/r)^2$. Boldyrev et al. [11] mentioned that the ion velocity does not change considerably the electric potential profile in the solar wind. Figure 3 illustrates the formation of an electron-trapping potential in our simulations. The results show that the EVDF for most electrons is close to isotropic, which justifies using a simpler model for these electrons.

Such a P1 model was implemented based on two-term Spherical Harmonics Expansion (SHE) of the EVDF in velocity space and coupled with the iterative solver for the self-consistent electric field.

Figure 4 shows the radial profiles of solution variables and the corresponding Electron Energy Distribution Functions (EEDF's) (an isotropic part of the EVDF) at two locations obtained from the P1 model.

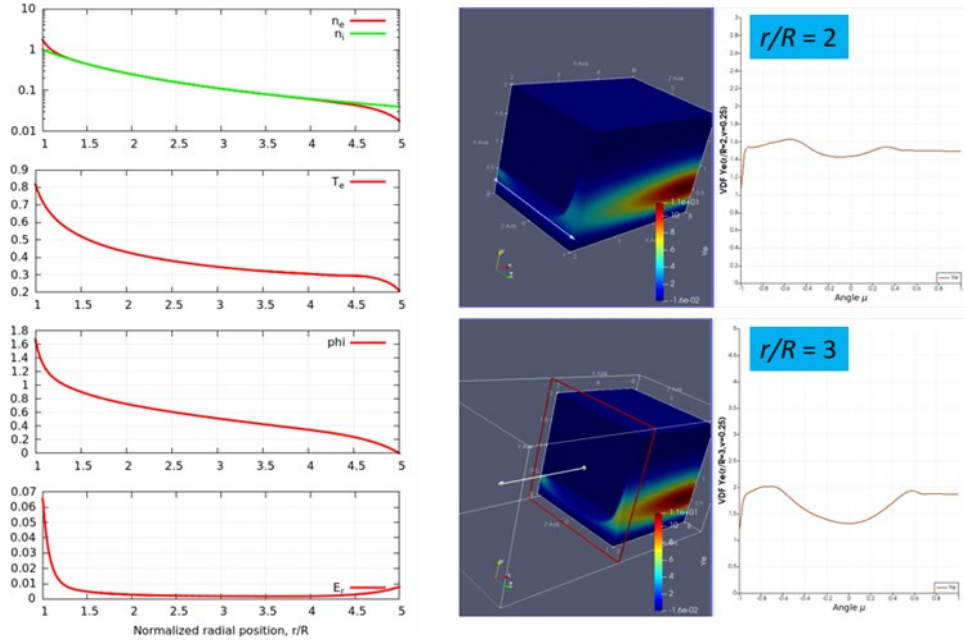

*Figure 3. 1D2V solutions for Debye length-to-radius ratio $\lambda_{De}/R = 0.1$. Shown are EVDF moments and self-consistent electrostatic potential and field (left) and EVDFs at two locations (right).*

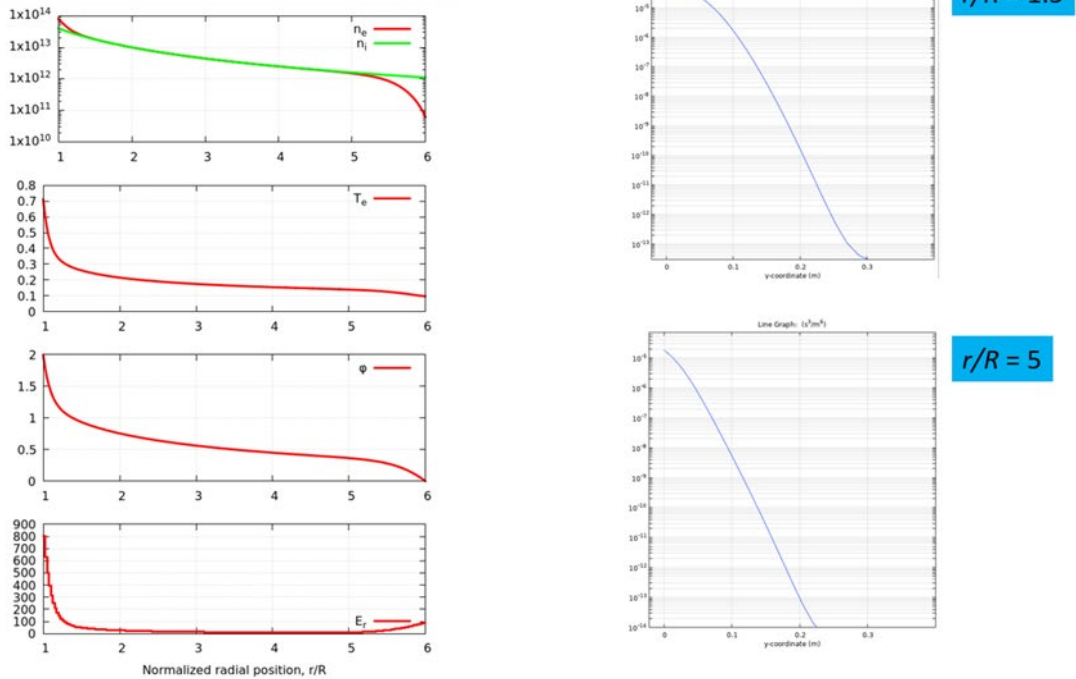

*Figure 4. The P1 solutions for Debye length-to-radius ratio $\lambda_{De}/R = 0.1$. EVDF moments and self-consistent electrostatic potential and field (left) and EEDFs at two locations (right).*

Figure 5 compares the results of 1D2V and P1 models. The trapping potential for electrons is formed in both cases. The decrease of electron temperature from the Sun's surface (predicted in Boldyrev et al.

[11]) is well reproduced by the P1-model code, with very similar (normalized) electron temperature magnitudes. The electrostatic potential and electric field predicted with the 1D2V and P1 models agree quantitatively.

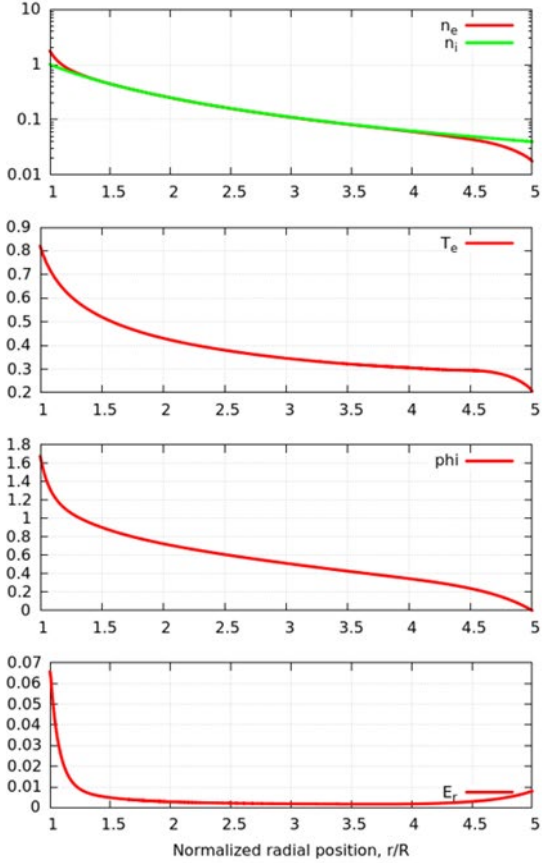 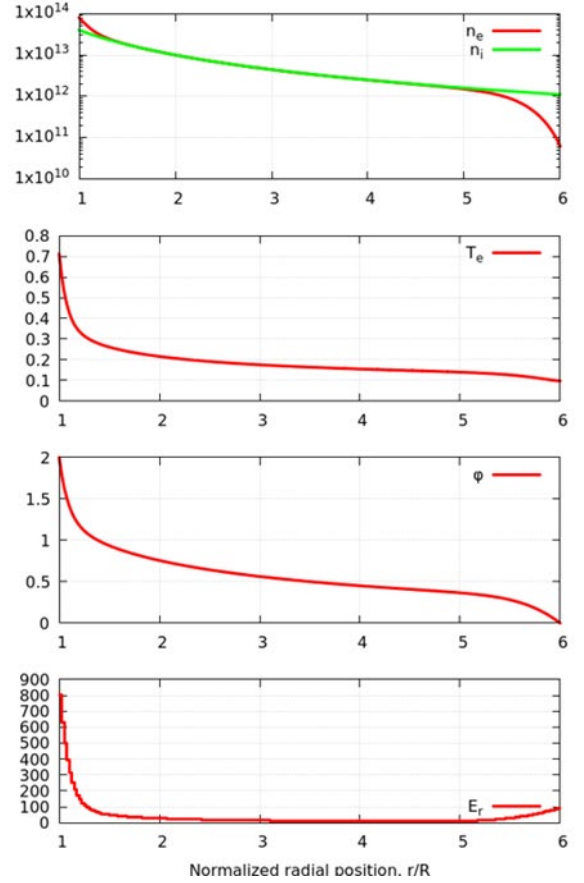

*Figure 5. The 1D2V (left) and P1-model (right) solutions for Debye length-to-radius ratio $\lambda_{De}/R = 0.1$.*

The overall agreement between the 1D2V model and the P1 model is encouraging. The P1 model can reproduce the main critical features of the VDF of returning electrons, which control the electric potential variation and magnitude at a significantly reduced cost compared to the 1D2V model. More detailed comparisons (e.g., for the particle and energy fluxes controlled by runaway anisotropic electrons) and model fine-tuning are required to validate the P1 model in a broader range of conditions.

To enable calculations for smaller Debye lengths, we implemented an iterative scheme for computations of (steady-state) electric potential based on the paper [15]. We used two schemes:

$$\varphi^{n+1}(r) = \varphi^n(r) + \alpha \varphi^{nocol}(r) \frac{n_e^n(r) - n_i^n(r)}{|j_e^n(r) + j_i^n(r)|} \qquad (4)$$

or

$$\varphi^{n+1}(r) = \varphi^n(r) + \alpha \varphi^{nocol}(r) \ln[n_e^n(r)/n_i^n(r)] \qquad (5)$$

Approximate analytic solutions for the plasma potential $\varphi^{nocol}(r)$ were applied as initial conditions to speed up convergence. The relaxation parameter $\alpha$ could be a function of $r$.

This Quasi-Neutral Field Solver (QNFS) allowed obtaining well-converged quasi-neutral solutions for low-resolution mesh in phase space without resolving space charge sheath. We have used a radially varying weight function for the relaxation parameter and verified that solutions are Debye-length independent. Figure 6 shows an example of calculations with a coarse static grid.

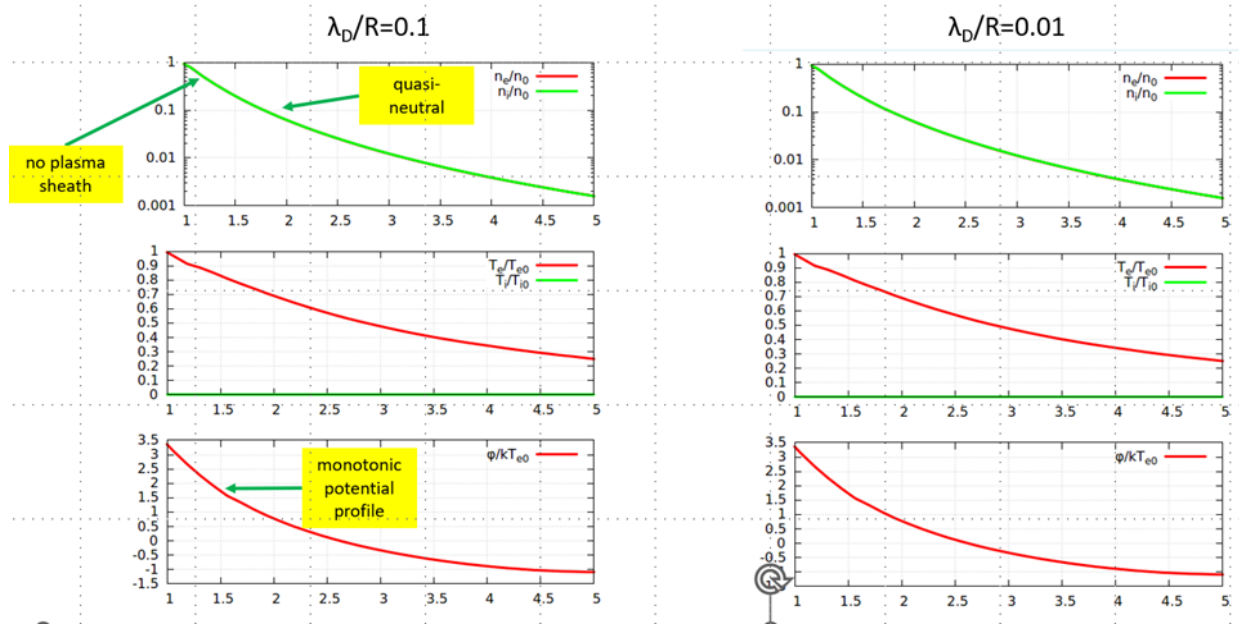

*Figure 6 Converged spatial distributions for different plasma densities (Debye lengths).*

We have enhanced QNFS for dynamically varying phase space mesh. Figure 7 shows a well-converged solution with Adaptive Mesh Refinement (AMR) in phase space (baseline grid + 1-level-up refinement). The dynamic AMR allows convergence on a coarse grid and then applies the iterative QNFS scheme. With this method, we first observed somewhat noisy solutions for the electric field, which were improved by implementing a noise-filtering technique. The QNFS module enables efficient computations under high plasma density conditions (extremely low Debye lengths), and we plan to expand the QNFS module further for simulations with ion transport and collisions.

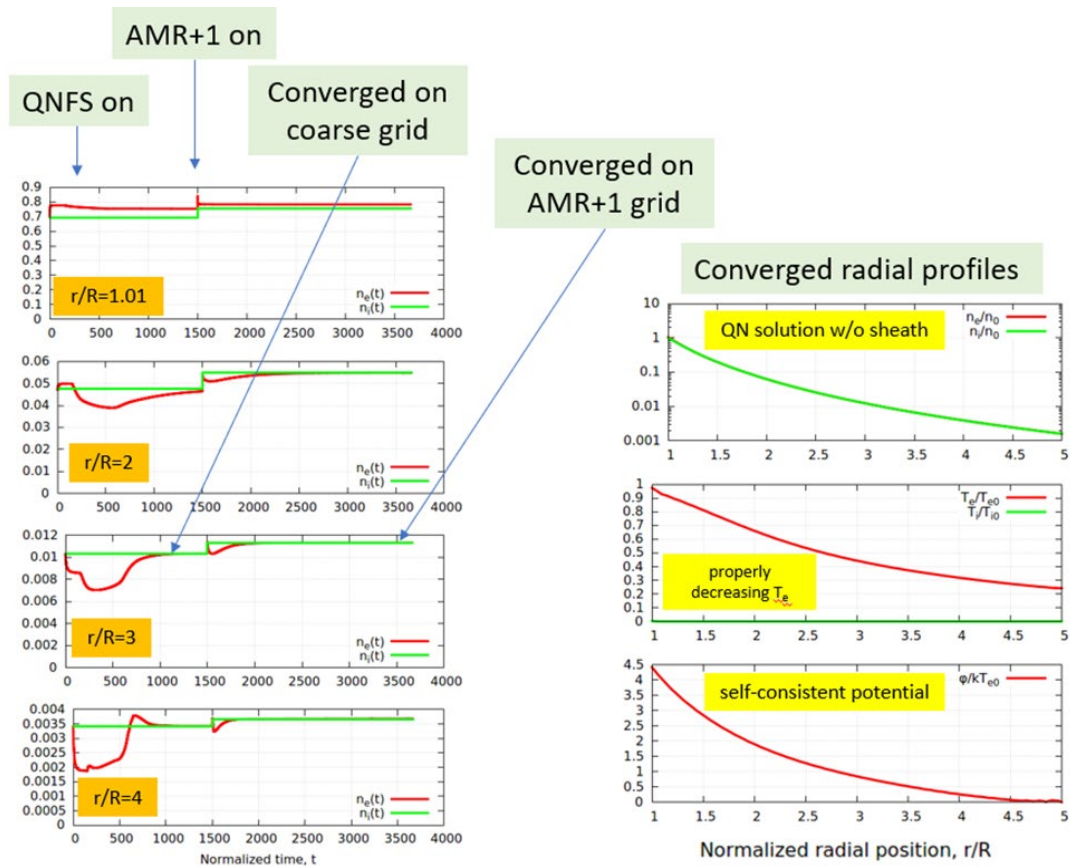

*Figure 7. Electron density convergence history with QNFS at different distances from the surface (left) and converged radial plasma profiles (right).*

**5. A quasi-diffusion model for electrons**
Quasi-diffusion models have been used for neutron, photon, and electron kinetics [16,17]. They have a broader applicability range than the P1 model described above, based on the two-term SHE. To evaluate the quasi-diffusion models proposed in [16], we have implemented a hybrid model of partially-ionized spherical plasma in COMSOL. The model uses a Fokker-Planck kinetic solver for EEDF, a free-flight fluid model for ions, and the Poisson equation for the electric field. Electrons and ions are injected from an internal sphere and exist through the external sphere. Elastic collisions of electrons with neutral gas atoms and Coulomb interactions among electrons are considered. Self-consistent distributions of the electric potential were found for different values of the potential drop between the spheres by solving a transient problem until a steady state. We observed that the electron flux in the steady state is reduced by varying the potential drop from 1.5 V to 3 V.

Below, we illustrate the model results for the spheres with a radius of 1 cm and 30 cm in Argon gas at a pressure of 1 mTorr. The electrons and ions are injected with densities $10^{18}$ cm$^{-3}$ and equal temperatures of 1eV, corresponding to the Debye length at the sphere location of 0.007 mm. Figure 8 shows the spatial distributions of the electron density and temperature and the EEDF at a specific spot for different values of the Coulomb collision frequencies at the assumed potential drop of 1.5 V between the spheres.

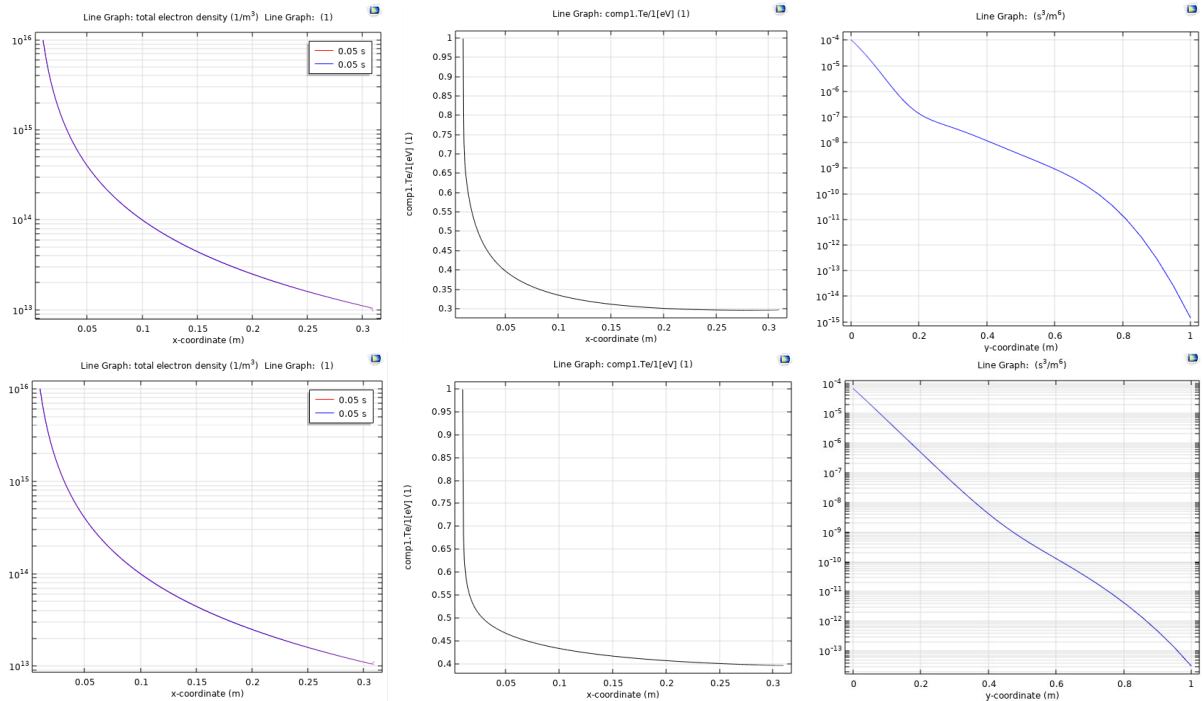

*Figure 8 The spatial distributions of the electron density and temperature and the EEDF at a specific location for different values of the Coulomb collision frequencies: $\nu_{ee} = 0$ (top) and $\nu_{ee} \neq 0$ (bottom).*

Figure 9 illustrates the effects of the Coulomb collisions on the spatial distributions of the electric potential. Minimal changes are observed when Coulomb collisions are taken into account.

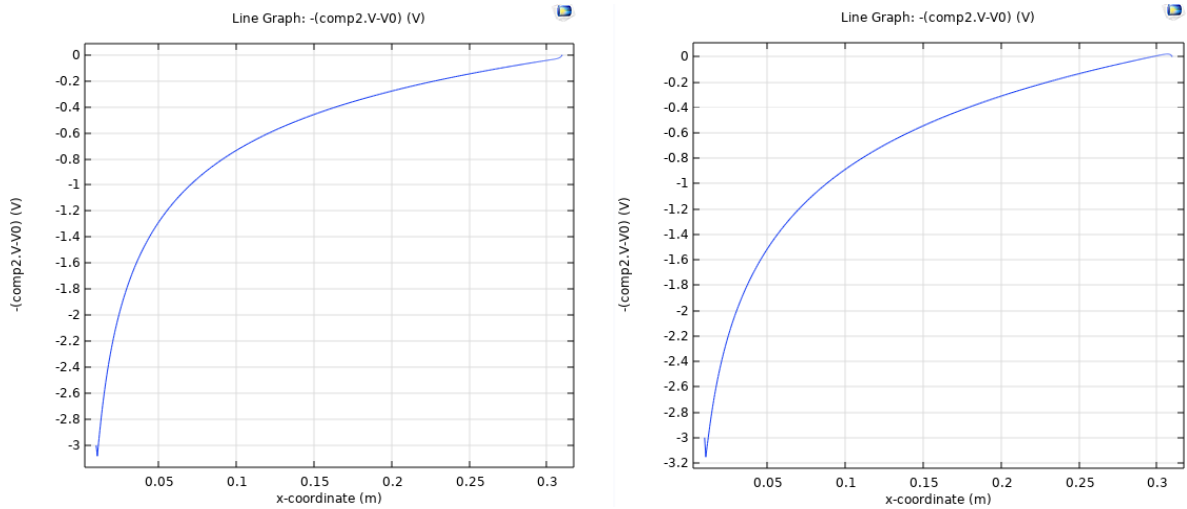

*Figure 9 The spatial distribution of the potential without (left) and with Coulomb collisions*

To summarize, partially-ionized spherical plasma with densities up to $10^{18}$ m$^{-3}$ could be modeled with COMSOL using a Poisson solver. For the considered test of the quasi-diffusion model, a potential well trapping most electrons formed and $T_e(r)$ dropped in a very similar fashion to full 1D2V results (from 1 to ~0.25). The EEDF of trapped electrons remained close to a Maxwellian (following the injected EEDF) even without Coulomb collisions—however, a significant "hot" tail with a temperature higher than the injected temperature formed. Coulomb collisions resulted in the transition to a Maxwellian EEDF with the same "temperature" for all electron energies, i.e., heating the cold, trapped electrons by the hot tail electrons. The spatial distributions of electron density $n(r)$ and $T_e(r)$ remained practically unchanged. This behavior indicates a proper numerical scheme, as Coulomb collisions among electrons should not affect the EEDF moments. We used a simple version of the quasi-diffusion model in these simulations, which corresponds to the P1 model used in the Basilisk simulations. We plan to combine the general quasi-diffusion model with the 1D2V solver for more efficient simulations of different electron groups.

**Acknowledgments**

This work was supported by the NSF EPSCoR project OIA-2148653 and NASA SBIR project 80NSSC-21-C-0527.